\documentclass[epj]{svjour}
\usepackage{graphicx,epsfig}
\usepackage{amsmath,amssymb}
\usepackage{array}
\usepackage{arydshln}
\usepackage{color}
\usepackage{cite}

\newcommand{\be}{\begin{equation}}
\newcommand{\ee}{\end{equation}}
\newcommand{\bea}{\begin{eqnarray}}
\newcommand{\eea}{\end{eqnarray}}
\newcommand{\beq}{\begin{equation}}
\newcommand{\eeq}{\end{equation}}
\newcommand{\beqa}{\begin{eqnarray}}
\newcommand{\eeqa}{\end{eqnarray}}
\newcommand{\ba}{\begin{array}}
\newcommand{\ea}{\end{array}}

\newcommand{\no}{\nonumber}

\newcommand{\cB}{{\cal B}}
\newcommand{\BR}{{\cal B}}
\newcommand{\cL}{{\cal L}}

\newcommand{\ord}[1]{O({#1})}

\definecolor{nicered}{rgb}{0.7,0.1,0.1}
\begin{document}

\title{Neutrino masses and LFV from minimal breaking of $U(3)^5$ and $U(2)^5$ flavor symmetries}

\author{
Gianluca Blankenburg\inst{1,2}, 
Gino Isidori\inst{2,3}, 
Joel Jones-P\'erez\inst{2,3}}

\institute{Dipartimento di Fisica, Universit\`a di Roma Tre,  Via della Vasca Navale 84, I-00146 Roma, Italy
\and
CERN, Theory Division, 1211 Geneva 23, Switzerland
\and
INFN, Laboratori Nazionali di Frascati, Via E.~Fermi 40, I-00044 Frascati, Italy}

\authorrunning{Blankenburg {\em et al.}}
\titlerunning{Neutrino masses and LFV from minimal breaking of $U(3)^5$ and $U(2)^5$ flavor symmetries}
\mail{gino.isidori@lnf.infn.it}

\date{Received: date / Revised version: date}

\abstract{We analyze neutrino masses and Lepton Flavor Violation (LFV) in charged leptons 
with a minimal ansatz about the breaking of the  $U(3)^5$ flavor symmetry, consistent with the 
$U(2)^3$ breaking pattern of quark Yukawa couplings, in the context of supersymmetry. 
Neutrino masses are expected to be almost degenerate, close to present bounds from cosmology 
and $0\nu\beta\beta$ experiments.
We also predict $s_{13} \approx  s_{23} |V_{td}|/|V_{ts}| \approx 0.16$, in perfect agreement
with the recent DayaBay result. For slepton masses below 1 TeV, 
barring accidental cancellations, 
we expect $\cB(\mu \to e \gamma) > 10^{-13}$ and $\cB(\tau \to \mu \gamma) > 10^{-9}$,
within the reach of future experimental searches.
}

\maketitle

\section{Introduction}

The Yukawa sector of the Standard Model (SM), 
minimally extended  with a Majorana mass terms for the left-handed neutrinos, provides an excellent 
description of all the observed phenomena of flavor mixing, both in the quark and in the lepton sector. 
Still, the peculiar pattern of quark and lepton masses 
seems to point toward some new dynamics responsible for this highly non-generic flavor pattern. 
Moreover, the instability of the Higgs sector under quantum corrections suggests the presence of 
physics beyond the SM close to the electroweak scale.

The absence of any deviation from the SM predictions in flavor-violating processes is one of the 
most severe problems in building realistic extensions of the model at the TeV scale. This is why minimal 
flavor-breaking hypotheses for SM extensions, such as the ansatz of Minimal Flavor Violation~\cite{MFV},
have been proposed. However, the MFV anastz does not provide any clue about the origin of the 
hierarchal nature of the Yukawas, and does not help in understanding the recent severe challenge 
to SM extensions posed by the absence of direct signals at the LHC. 

A TeV extension of the SM aimed to address, at least in part, both the stability of the electroweak sector 
and the flavor problem is supersymmetry with heavy squark masses for the first two 
families of squarks, in short {\em split-family} SUSY.  While the stabilization of the Higgs sector  
requires mostly the third generation squarks to be light, the tight constraints from CP- and flavor-violating
processes are loosened in presence of a squark mass hierarchy~\cite{Dimopoulos:1995mi}.
In addition, while the bounds on first generation squark masses are already exceeding 1~TeV, 
the third generation squarks can still be significantly lighter~\cite{Papucci:2011wy}.
As pointed out in a series of recent works~\cite{Craig:2011yk},
 such a split spectrum can be achieved with realistic ultraviolet completions of the model.

A hierarchical squark spectrum is not enough to suppress flavor violation to a level consistent with experiments.
This is why split-family SUSY with a minimally broken $U(2)^3=U(2)_q \times U(2)_d \times U(2)_u$ 
flavor symmetry, acting on the first two generations of quarks (and squarks),  has been considered in Ref.~\cite{Barbieri:2011ci}.
This set-up has the following advantages:  i) it provides some insights about the structures of the Yukawa couplings
(along the lines of  $U(2)$ models proposed long ago~\cite{U2}); 
ii) it ensures a sufficient protection of flavor-changing neutral currents; iii) it leads to an improved 
CKM fit with tiny and correlated non-standard contributions to $\Delta F=2$ observables. 
Possible signatures of this framework in the $\Delta F=1$ sector have been discussed in Ref.~\cite{Barbieri:2011fc} (see 
also Ref.~\cite{Crivellin:2011fb}, where the same symmetry with additional dynamical assumption has been considered).
More general discussions about the  $U(2)^3$ flavor symmetry beyond supersymmetry has 
recently been presented in Ref.~\cite{Barbieri:2012uh,Redinew}.

The purpose of this article is to extend the idea of a minimally broken flavor symmetry acting on the first two 
generations to the lepton sector. The extension is straightforward in the case of  charged leptons, enlarging 
the flavor symmetry from $U(2)^3$ to $U(2)^5=U(2)^3\times U(2)_l \times U(2)_e$. However, the situation is more
involved in the neutrino sector, whose mass matrix has a rather different flavor structure: no large 
hierarchies in the eigenvalues, and large mixing angles~\cite{Glashow:2011bz}. A simple ansatz to circumvent this problem 
is to a assume a two-step breaking in the neutrino sector: first, a leading breaking of the maximal flavor symmetry, 
$U(3)_l \times U(3)_e$, that includes the total lepton number (LN), 
giving rise to a fully degenerate neutrino spectrum. This would be followed by a sub-leading LN-conserving 
breaking with a hierarchical structure similar to the one occurring in the charged-lepton sector. 
As we discuss, this minimal breaking structure gives rise to a phenomenologically viable neutrino mass matrix,
with a few interesting predictions concerning $s_{13}$ and the overall scale of neutrino masses.
It also predicts LFV in charged leptons compatible with present bounds and not far from the sensitivity of 
 future experimental searches in the case of $\mu \to e \gamma$ and $\tau \to \mu \gamma$.

The paper is organized as follows: in Section~\ref{sec:mod.ind} we discuss the general structure 
of the lepton mass matrices implied by present data, trying to identify possible ``starting points" 
for the neutrino mass matrix in the limit of unbroken flavor symmetry.
In Section~\ref{sec:nu.model} we present our hypotheses about the (lepton) flavor symmetry and 
its breaking pattern.  The phenomenological consequences of this ansaz for the neutrino mass
matrix and LFV are discussed in Section~\ref{sec:pred.nu} and \ref{sec:pred.lfv}, 
respectively. The results are summarized in the Conclusions.

\section{General considerations on lepton masses}
\label{sec:mod.ind}
We define the charged lepton Yukawa coupling ($Y_e$) and the effective neutrino 
Majorana mass matrix ($m_\nu$) from the following effective Lagrangian, written in terms of 
SM fields:
\be
\cL^{\rm eff}_{\rm mass} = \bar L_{Li} (Y_e^*)_{ij} e_{Rj} H
+ (m_\nu)_{ij} \,\bar\nu^{ci}_L\nu^j_L + {\rm h.c.}
\ee
As usual, the neutrino mass term can interpreted as the result 
of an appropriate dimension-five gauge invariant  
operator after the spontaneous breaking of the electroweak symmetry~\cite{Weinberg:1979sa}, 
with the Higgs vacuum expectation value absorbed in the effective coupling $m_\nu$.

By construction, $M^2_\nu= m_\nu^\dagger m_\nu$ and $Y^*_e Y^T_e$ 
transform in the same way under flavor rotations of the left-handed lepton doublets,
while they are invariant under rotations in the right-handed sector. 
The charged lepton sector exhibits a strongly hierarchical structure: with a proper basis choice 
for the left-handed fields,
and neglecting entries of $O(m_\mu/m_\tau)$, we have
\be
 Y^*_e Y^T_e \approx  (Y^*_e Y^T_e)^{(0)}  = y^2_\tau ~ {\rm diag}(0,0,1)~,
 \label{eq:Ye0}
\ee 
where $y_\tau = \sqrt{2} m_\tau/v$ ($v\approx 246$~GeV).

In the basis where $Y_e$ is diagonal, $M^2_\nu$ assumes the form
\beq
M^2_\nu  = U_{\rm PMNS} (m^2_{\nu})^{\rm diag} U_{\rm PMNS}^\dagger~,
\eeq
where $U_{\rm PMNS}$ is the so-called PMNS matrix.
We adopt the PDG parameterization~\cite{PDG}, such that the mass eigenstates are ordered following a normal hierarchy ($m_{\nu_1} < m_{\nu_2} < m_{\nu_3}$) or an inverted one ($m_{\nu_3} <  m_{\nu_1}  <  m_{\nu_2}$). To distinguish between them, one defines $\Delta m^2_{ij}=m^2_{\nu_i}-m^2_{\nu_j}$, such that $\Delta m^2_{31}= \pm  \Delta m^2_{\rm atm}$ and 
$\Delta m^2_{21} =  \Delta m^2_{\rm sol}$, where $\Delta m^2_{\rm atm, sol}$ denote the  (positive) 
squ\-ared mass differences deduced from atmospheric and solar neutrino data.
It is straightforward to deduce that the plus (minus) sign of $\Delta m^2_{31}$ corresponds to the normal (inverted) hierarchy.

Experimental data on neutrino oscillations indicate the presence of (at least) two small parameters
in $M^2_\nu$,
\begin{eqnarray}
\label{eq:smallp}
 \zeta &=& \left| \frac{\Delta m^2_{\rm sol}}{\Delta m^2_{\rm atm}} \right|^{1/2}~,\qquad  \zeta^{\rm exp}= 0.174\pm 0.007~, \label{eq:zeta} \\
 s_{13} &=& \left| (U_{\rm PMNS})_{13} \right|~, \qquad s_{13}^{\rm exp}  = 0.15\pm 0.02~,  \label{eq:s13exp}
\end{eqnarray}
where the value of  $s_{13}$ has been determined from the recent result of the DayaBay experiment~\cite{An:2012eh}.
Expanding to lowest order in these two parameters 
(or in the limit $\zeta, s_{13} \to 0$) 
 we are left with the following structure 
 \begin{equation}\
\label{eq-start1}
(M_\nu^2)^{(0)} =  m_{\textrm{light}}^2 \cdot I + \Delta m^2_{\rm atm} \cdot \Delta
\ee
where $I$ is the identity matrix, $m_{\textrm{light}}$ is the lightest neutrino mass, and 
\bea
\Delta_{[\textrm{n.h.}]} &=&  \left(
\begin{array}{ccc}
0 & 0 & 0\\0 & s_{23}^2 & s_{23}c_{23}\\0 & s_{23}c_{23} & c_{23}^2 
\end{array}\right) \approx \frac{1}{2} \left( \begin{array}{ccc} 0 & 0 & 0\\0 & 1 & 1\\0 & 1 & 1  \end{array}\right)~, \label{eq:DeltaR} \\
\Delta_{[\textrm{i.h.}]} &=&  I -  \Delta_{[\textrm{n.h.}]} ~.  \label{eq:DeltaR2}
\eea
In order to define a starting point for the neutrino mass matrix in the limit of unbroken flavor 
symmetry we need to specify the hierarchy between $m^2_{\textrm{light}}$ and $\Delta m^2_{\rm atm}$,
or among the two terms in Eq.~(\ref{eq-start1}). We thus have three natural possibilities:
\begin{itemize}
\item[I.]  $(M_\nu^2)^{(0)} \propto   I$, if  $m^2_{\textrm{light}} \gg \Delta m^2_{\rm atm}$,
\item[II.] $(M_\nu^2)^{(0)} \propto   \Delta_{[\textrm{n.h.}]}$, if  $m^2_{\textrm{light}} \ll \Delta m^2_{\rm atm}$ and $\Delta m^2_{31}>1$,
\item[III.] $(M_\nu^2)^{(0)} \propto   \Delta_{[\textrm{i.h.}]}$, if  $m^2_{\textrm{light}} \ll \Delta m^2_{\rm atm}$ and $\Delta m^2_{31}<1$.
\end{itemize}

\section{Flavor symmetries  and symmetry  breaking}
\label{sec:nu.model}

\subsection{$U(2)_l \times U(2)_e$}
The $U(2)^2 = U(2)_l \times U(2)_e$ flavor symmetry, under which the lepton superfields of the first two 
families transform as 
\begin{eqnarray}
  L_L  \equiv ( L_{L1}  , L_{L2} )^{\phantom{T}}  &\sim& (\bar 2,1)~, \\
  e^{c}\equiv ( e_{1}^c , e_{2}^c )^T    &\sim& (1,2)~, 
\end{eqnarray}
offers a natural framework to justify the hierarchal 
structure of the charged-lepton Yukawa coupling, in close analogy 
to the $U(2)^3$ symmetry introduced in Ref.~\cite{Barbieri:2011ci} for the quark sector.
In the limit of unbroken symmetry we recover the result in Eq.~(\ref{eq:Ye0}).
Assuming a symmetry-breaking 
pattern  for $Y_e$ similar to the one adopted for the quark Yukawa couplings, we get
\be
\label{eq:Yefull}
Y_e = y_\tau \left(\begin{array}{c:c}
 \Delta Y_e & V \\\hdashline
 0 & 1
\end{array}\right),
\ee
where we have absorbed $\ord{1}$ couplings in the definition of the breaking terms $V\sim(2,1)$ 
and $\Delta Y_e \sim (2,\bar 2)$.  Introducing the unitary matrices $U_{eL}$ and $U_{eR}$,
such that 
\be
 U_{eL} Y_e U_{eR}^\dagger  =  {\rm diag}(y_e,y_\mu,y_\tau)~, 
\ee
and proceeding as in~\cite{Barbieri:2011ci}, we find that $U_{eR}$ becomes the identity matrix 
in the limit $m_{e,\mu}/m_\tau \to 0$, while $U_{eL}$ assumes the following parametric form
\be
\label{eq:lepmix} 
U_{eL}\approx\left(\begin{array}{ccc}
c_e & s_e\,c_{\tau}\,e^{i\alpha_e} & -s_e\,s_{\tau}\,e^{i(\alpha_e+\phi_\tau)} \\
-s_e\,e^{-i\alpha_e} & c_e\,c_{\tau} & -c_e\,s_{\tau}\,e^{i\phi_\tau} \\
0 & s_{\tau}\,e^{-i\phi_\tau} & c_{\tau}
\end{array}\right)~,
\end{equation}
in the $U(2)_l$ basis where $V^T\propto (0,1)$. Here 
  $\alpha_e$ and $\phi_\tau$ are generic $O(1)$ phases, while 
$s_{e,\tau}$ are small mixing angles ($c^2_{i}+s^2_{i}=1$).
If the analogy with the quark sector holds, we expect $s_e$ to be 
of the order of $s_d=|V_{td}|/|V_{ts}|\approx 0.22$ and $s_{\tau}$ of the order of 
$\epsilon=|V_{cb}|\approx 0.04$.

From the point of view of the $U(2)^2$ symmetry, the neutrino mass matrix can be decomposed as 
\begin{equation}\label{mvu2}
m_\nu = \left(
\begin{array}{c:c}
m_3 & m_2\\ \hdashline m_2^T & m_1 
\end{array}\right)~,
\end{equation}
where $m_3\sim (3,1)$,  $m_2\sim (2,1)$, and $m_1\sim (1,1)$. This decomposition does not match well
with any of the potential starting points identified in Eqs.~(\ref{eq-start1})--(\ref{eq:DeltaR2}): they 
can be obtained only assuming specific relations among terms with different $U(2)_l$
transformation properties. This suggests that we need to consider a larger flavor symmetry,
whose breaking to $U(2)_l$ (or some of its subgroups) could explain such relations.
From this point of view the degenerate case is the one that 
offers the most interesting prospects: on the one hand it requires a special relation only 
among two of the terms appearing in Eq.~(\ref{mvu2}): $m_3={\rm diag}(m_1,m_1)$.
On the other hand, it requires $m_2 \ll 1$, as expected given that $m_2$ transforms
as the breaking spurion $V$ of $O(\epsilon)$ appearing in the charged-lepton Yukawa coupling.  
As we discuss in the following, the degenerate case can easily be obtained embedding 
$U(2)_l$ in $U(3)_l$.

\subsection{$U(3)_l \times U(3)_e$}
The group $U(3)_l \times U(3)_e$ is the largest flavor symmetry of the lepton sector allowed by the SM gauge Lagrangian.
The degenerate configuration for $m_\nu$ is achieved assuming that  $U(3)_l$ and the total 
lepton number, 
\be
U(1)_{\rm LN}=U(1)_{l+e}~, 
\ee
are broken by a spurion $m_\nu^{(0)}$ transforming\footnote{~We denote in bold $U(3)$ vectors and representations.} as a ${\mathbf 6}$ of $U(3)_l$ and leaving invariant a subgroup of $U(3)_l$ that we denote $O(3)_l$.
By a proper basis choice in the $U(3)_l$ flavor space we can set 
\be
m_\nu^{(0)} \propto \left(\begin{array}{c:c}
 I & 0 \\ \hdashline 0 & 1
 \end{array}\right)~.
\ee

We shall also require that $U(3)_l \times U(3)_e$ is broken by $U(1)_{\rm LN}$ invariant spurions
to the subgroup $U(2)_l \times U(2)_e$ relevant to the charged-lepton Yukawa coupling.
However, it is essential
for our construction that this (sizable) breaking  does not spoil the Majorana sector, at least in first approximation. 
This can be achieved in a supersymmetric context introducing a new spurion $Y^{(0)} \sim ({\mathbf 3}, \bar {\mathbf 3})$
that breaks $U(3)_l\times U(3)_e$ to $U(2)_l\times U(2)_e$ leaving unbroken the $O(2)_l$ subgroup of both 
of  $O(3)_l$ and $U(2)_l$.
By means of $Y^{(0)}$ we can have a non-vanishing Yukawa coupling for the third generation 
in the superpotential
\be
{\mathbf L_L} Y^{(0)} {\mathbf e^c} \to  y^{(0)}_\tau  L_3  e^c_3~.
\label{eq:Y0}
\ee
and, in first approximation, the Majorana mass matrix is unchanged. 
Note that supersymmetry is a key ingredient for the the latter statement to hold.  
Indeed, if the mass operator was not holomorphic, a Majorana term of the type
${\mathbf L_L} Y\,Y^\dag\, m_\nu^{(0)}\, {\mathbf L_L^T}$ could also be included and this would spoil the degenerate configuration.

Summarizing, introducing the two spurions $m_\nu^{(0)}$ and  $Y^{(0)}$ we recover phenomenologically viable first approximations to 
both the neutrino and the charged-lepton mass matrices and we are left with an exact $O(2)_l \times U(2)_e$ symmetry
that leaves invariant both $m_\nu$ and  $Y_e$. Moreover, thanks to supersymmetry, the two sector considered separately are
invariant under larger symmetries: $O(3)_l$ for the neutrinos and $U(2)_l\times U(2)_e$ for the charged leptons.

We can then proceed introducing the small $O(2)_l \times U(2)_e$ breaking terms responsible for the subleading terms in $Y_e$
in Eq.~(\ref{eq:Yefull}).   In order not to spoil the leading structure of the neutrino mass matrix,
the spurion $V$ in Eq.~(\ref{eq:Yefull}) should be regarded as a doublet of $O(2)_l$, rather than a doublet of $U(2)_l$.\footnote{~This has 
no practical implications if we consider the Yukawa sector alone, where there in no preferred $O(2)$ subgroup of $U(2)_l$
in the limit $V\to 0$.}
We can also regard it as the $O(2)_l$ component of an appropriate ${\mathbf 8}$ of  $U(3)_l$ with the following structure
\be
X= \left(
\begin{array}{c:c}
\Delta_L & V \\ \hdashline V^\dagger & x 
\end{array}\right)~.
\ee
This allows to write the additional Yukawa interaction ${\mathbf L_L} X\,Y^{(0)}\, {\mathbf e^c}$ that, 
combined with the leading term in (\ref{eq:Y0}) and with a proper redefinition of $y_\tau$ and $V$ 
implies
\be
Y^{(1)}_e = y_\tau\,\left(
\begin{array}{c:c}
0 & V \\ \hdashline 0 & 1
\end{array}\right)~.
\ee
All the components of $X$ do appear in the Majorana sector, via the terms
${\mathbf L_L} X\, m_\nu^{(0)}\, {\mathbf L_L^T}$
and ${\mathbf L_L} m_\nu^{(0)} \,X^T {\mathbf L_L^T}$. These imply the following structure 
\begin{equation}
m_\nu= m_{\nu_1}^{(0)} \left[ I + a\,  \left( 
\begin{array}{c:c}
 \Delta_L & V \\ \hdashline  V^T &  x 
\end{array}\right) \right]~,
\end{equation}
where $a$ is a  $O(1)$ complex coupling. Assuming that all the entries of $X$ are at most of $O(\epsilon)$
does not spoil the degenerate configuration of $m_\nu$ in first approximation. In addition, since 
$\Delta_L$ could enter linearly in the sfermion mass matrices and induce sizable FCNCs, 
we expect a small mis-alignement between $\Delta_L$ and $V$ in the $O(2)_l$ space. Pursuing the analogy with the squark sector, we are 
forced to assume $(\Delta_L)_{12}$ at most of $O(\epsilon^2)$ in the basis where $V_1=0$.
In other words, we are lead to the following assignment for the various components of $X$ in the $O(2)_l$ basis 
where $V^T\propto(0,1)$:
\begin{equation}
V= \left( \begin{array}{c} 0 \\  O(\epsilon) \end{array} \right) , \quad 
\Delta_L=\left(\begin{array}{cc}
0 & O(\epsilon^2) \\
O(\epsilon^2) & O(\epsilon)\\
\end{array}\right), \quad x=O(\epsilon).
\end{equation}
In the same basis, redefining the unknown parameters, we arrive to the following parametric expression
\be
\label{mvok}
m_\nu= \bar m_{\nu_1}\left[ I + e^{i\phi_\nu} \left(
\begin{array}{ccc}
-\sigma \epsilon & \ \gamma\epsilon^2 &  0 \\
\ \gamma\epsilon^2 & -\delta \epsilon &  \  r\epsilon \\
0 & \ r\epsilon & 0
\end{array}\right)\right]~, 
\ee
where $\phi_\nu$, $\sigma$, $\delta$, $\gamma$, and $r$ are 
real parameters expected to be of $O(1)$.

The final step for the construction of a realistic charged-lepton Yukawa coupling 
is the introduction of the $U(2)_l \times U(2)_e$ bi-doublet $\Delta Y_e$. The most economical way 
to achieve this goal in the context of $U(3)_l \times U(3)_e$ is to introduce a bi-triplet 
with the following from,
\be
\Delta \hat Y_e = \left(
\begin{array}{c:c}
\Delta Y_e & 0 \\ \hdashline 0 & 0
\end{array}\right)~,
\ee
which provides the desired correction to $Y_e$ and has no relevant 
impact on $m_\nu$.

Notice that the requirement of having $Y^{(0)}$ and $X$ acting on $O(3)$ and $O(2)$ subspaces can be naturally accomplished by demanding an exact CP symmetry acting on both spurions. The CP symmetry would be broken only by the $\Delta \hat Y_f$ spurions, which would provide all CP violation phases.\footnote{For quarks, it was shown in~\cite{Barbieri:2011ci} that the CKM phase is entirely defined by the phases in the $\Delta Y_f$ spurions. Thus, the conclusions for the quark sector would remain unchanged.}

Our parametrical decomposition of the neutrino mass matrix is therefore the expression in Eq.~(\ref{mvok}). 
As can be seen, the latter contains only a CP-violating phase, $\phi_\nu$, which does not contribute 
to the PMNS matrix. However,  this does not imply that there are no CP-violating 
phases in $U_{\rm PMNS}$: a non-vani\-shing phase arises by the diagonalization of the charged-lepton mass matrix. 
Indeed, to leading order in $\epsilon$, the parametric decomposition of 
$M^2_\nu =m_\nu^\dagger m_\nu$ in the basis where $Y_e$ is diagonal is 
\be
 M^2_\nu = \bar m^2_{\nu_1}   U^T_{eL} \left(\begin{array}{ccc}
 1-2\sigma\,\epsilon &  & \\
 2\gamma\,\epsilon^2 & 1-2\delta\epsilon & \\
 O(\epsilon^3)  & 2r\,\epsilon & 1
\end{array}\right)  U^*_{eL}~,\label{mat:mass}
\end{equation}
where we have redefined  $\sigma$, $\delta$, $\gamma$, and $r$, absorbing a $\cos(\phi_\nu)$ term
and $U_{eL}$ is given in Eq.~(\ref{eq:lepmix}).
Despite the presence of  four O(1) free parameters, the expression of  $M^2_\nu$ in Eq.~(\ref{mat:mass})
 is quite constrained by  
the smallness of $\epsilon$. As we discuss in the next section, it provides a good fit to all the available neutrino data
for a natural range of the free parameters and leads to a few unambiguous predictions. We finally stress that we arrived 
to this decomposition using essentially two main assumptions:
\begin{itemize}
\item[I.] An approximate degenerate neutrino spectrum, that fits well with present data if $m^2_{\textrm{light}} \gg \Delta m^2_{\rm atm}$.
\item[II.] A symmetry-breaking pattern with respect to a purely degenerate spectrum closely related to
the minimal $U(2)^5$ symmetry breaking pattern of quark and lepton Yukawa couplings.
\end{itemize}

\section{Predictions for neutrino masses and mixings}
\label{sec:pred.nu}

We are now ready to analyze the predictions of the $M^2_\nu$  parameterization 
in Eq.~(\ref{mat:mass}). We first discuss a few simple analytic results, valid to leading order in $\epsilon$. Given the 
neutrino spectrum is almost degenerate, the ordering of the eigenvalues has 
 no physical implications. However, for the sake of simplicity, we present analytic results 
only in the case of normal ordering  ($m_{\nu_3} > m_{\nu_2} > m_{\nu_1}$).  
 We then proceed with a systematic numerical scan of the four O(1) 
free parameters to investigate the stability of the analytic conclusions.

\subsection{Mass eigenvalues}

From the decomposition of  $M^2_\nu$ in Eq.~(\ref{mat:mass}) we derive the following 
expressions for the eigenvalues, 
\begin{eqnarray}
 m_{\nu_1}^2 &=& \bar m^2_{\nu_1}\left(1-2\sigma\,\epsilon\right)~, \\
 m_{\nu_2}^2 &=& \bar m^2_{\nu_1}\left[1-\delta\,\epsilon-(\delta^2+4r^2)^{1/2}\epsilon\right]~, \\
 m_{\nu_3}^2 &=& \bar m^2_{\nu_1}\left[1-\delta\,\epsilon+(\delta^2+4r^2)^{1/2}\epsilon\right]~,
\end{eqnarray}
up to $O(\epsilon^2)$ corrections.  The normal ordering of the spectrum is obtained for 
$\sigma > (\delta^2+4r^2)^{1/2}$ and in this case we find
\bea
\frac{ \Delta m^2_{\rm atm} }{m^2_{\nu_1}} &=& \left[ 2\sigma-\delta+(\delta^2+4r^2)^{1/2}\right] \epsilon~, 
\label{eq:Datm} \\
 \zeta  &=&   \left[ \frac{2\sigma-\delta-(\delta^2+4r^2)^{1/2}}{2\sigma-\delta+(\delta^2+4r^2)^{1/2}} \right]^{1/2}~.
\eea
As can be seen, $\epsilon$ controls the overall scale of neutrino masses, whose natural scale is 
\be
 O[ (\Delta m^2_{\rm atm})^{1/2} \epsilon^{-1/2} ] = O(0.3~{\rm eV}),
\ee
just below the existing bounds (see discussion at the end of this Section).
Our parametric decomposition of $M^2_\nu$ does not necessarily imply 
$\zeta \ll 1$. However, the experimental value in Eq.~(\ref{eq:zeta}) is easily obtained 
with a modest tuning of the free parameters, especially if $\delta$ is small.
As we discuss next,  the latter is a condition necessary to reproduce the maximal 2-3
mixing in the PMNS matrix. 

\subsection{$\theta_{23}$}

In order to determine the $\theta_{23}$ and $\theta_{13}$ mixing angles of the 
PMNS matrix it is sufficient to expand $M^2_\nu$  up to $O(\epsilon^2,s_e^2\epsilon)$:
\begin{equation}
 M^2_\nu=\bar m_{\nu_1}\left(\begin{array}{ccc}
 1-2\epsilon \sigma & & \\
-2 s_e \epsilon(\sigma-\delta)\,e^{i\alpha_e} & 1-2\epsilon \delta & \\
-2\epsilon\,s_e r \,e^{i\alpha_e} & 2\epsilon r & 1
\end{array}\right) +O(\epsilon^2,s_e^2\epsilon)~. \label{eq:MNexp1} \\
\end{equation}
As can be seen, at this level $\gamma$ does not appear. 
Note also that for $s_e \to 0$ the 2-3 sector decouples. In this limit we obtain 
the following simple expression for the 2-3 mixing:
\begin{equation}
\label{eq:t23}
 t_{23} = \frac{s_{23}}{c_{23}}  = \frac{\delta \pm [\delta^2+4r^2]^{1/2}}{2r}~,
\end{equation}
where the sign in the numerator is chosen depending on the sign of $r$, such that $t_{23}$ remains positive. 
As we have explicitly checked by means of the numerical scan, this result is stable with respect to the inclusion 
of the subleading terms of $O(s_e \epsilon)$ and $O(\epsilon^2)$. 

From Eq.~(\ref{eq:t23}) it is clear that  $t_{23}$ is naturally expected to be $O(1)$, and the 
experimental evidence of maximal mixing, $t_{23}\approx1$, is obtained for $\delta \to 0$.

\subsection{$\theta_{13}$ and the PMNS phase}

The value of $\theta_{13}$ can be obtained via the following general relation 
\be
\frac{(M^2_\nu)_{31}}{(M^2_\nu)_{32}} \approx \frac{s_{13}}{s_{23}}\,e^{i\delta_{\rm P}}~,
\label{pred2}
\ee
that can be derived expanding $M^2_\nu$ --in the basis where $Y_e$ is diagonal--
up to the first order in $\zeta$ and $s_{13}$ (here $\delta_{\rm P}$ denotes the PMNS phase 
in the standard parameterization). Applying this result to the approximate from in  Eq.~(\ref{eq:MNexp1})
leads to 
\begin{equation}
\label{eq:s13}
s_{13} e^{i \delta_P }  = s_e s_{23} e^{\alpha_e+\pi}~.
\end{equation}
Assuming $s_e = s_d =|V_{td}|/|V_{ts}|$, and the experimental value of 
$s_{23}$ ($s^2_{23} = 0.52 \pm 0.06$~\cite{Schwetz:2011zk}), we predict
\be
s_{13} = 0.16 \pm 0.02~,
\ee
in remarkable agreement with the recent DayaBay result in Eq.~(\ref{eq:s13exp}).
This prediction is affected by a theoretical error (not explicitly shown) 
due to possible deviations from the relation $s_e = s_d$. On the other hand, 
as we have checked by means of the numerical scan, it is quite stable 
with respect to subleading corrections in the expansion of $M^2_\nu$. 

As anticipated, the PMNS phase is completely determined in terms of 
the CP-violating phase from the rotation of the charged-lepton Yukawa  
coupling: $\delta_P=\alpha_e+\pi$.  However, we are not able to determine 
this phase even assuming $\alpha_e=\alpha_d$, where $\alpha_d$ is 
the corresponding  phase appearing in the diagonalization 
of $Y_d$.\footnote{The physical phase appearing in the CKM matrix 
can be determined in terms of $\alpha_d-\alpha_u$~\cite{Barbieri:2011ci}, but we
cannot disentangle  $\alpha_d$ and $\alpha_u$ without extra theoretical assumptions.}
On general grounds, we expect $\delta_P$ to be a generic $O(1)$ phase.
 
\subsection{$\theta_{12}$}

Contrary to the case of $\theta_{23}$ and $\theta_{13}$, the determination of  $\theta_{12}$ involve subleading terms
in $M^2_\nu$ and thus is more unstable. 

As an illustration, consider $M^2_\nu$ in the limit $s_e\rightarrow0$.  In this simplified 
case we obtain the relation 
\begin{equation}
 \tan2\theta_{12}=\frac{2\gamma\,c_{23}}{\sigma-\delta\,c^2_{23}-2r\,s_{23}c_{23}}\epsilon~,
\end{equation}
that seems to imply $\theta_{12} \approx 0, \pi/2$. However, once we impose the constraints from the squared mass differences, 
we find a cancellation in the denominator leading to generic $O(1)$ values for $\theta_{12}$. More explicitly,
expressing $s_{23}$ and $c_{23}$ in terms of $\delta$ and $r$ by means of Eq.~(\ref{eq:t23}) we find 
$$
\tan2\theta_{12} = \frac{4\gamma\,\epsilon}{2\sigma-\delta-[\delta^2+4r^2]^{1/2}}c_{23} =O(1) \times \frac{\epsilon}{\zeta^2}~,
$$
which is manifestly a generic O(1) number. This general conclusion remains valid when subleasing terms of 
$O(s_e \epsilon)$ and $O(\epsilon^2)$ are taken into account, although the explicit analytical expression for $\theta_{12}$
becomes more involved.

\subsection{Parameter Scan}

\begin{figure}[t]
\begin{center}
\includegraphics[width=0.4\textwidth]{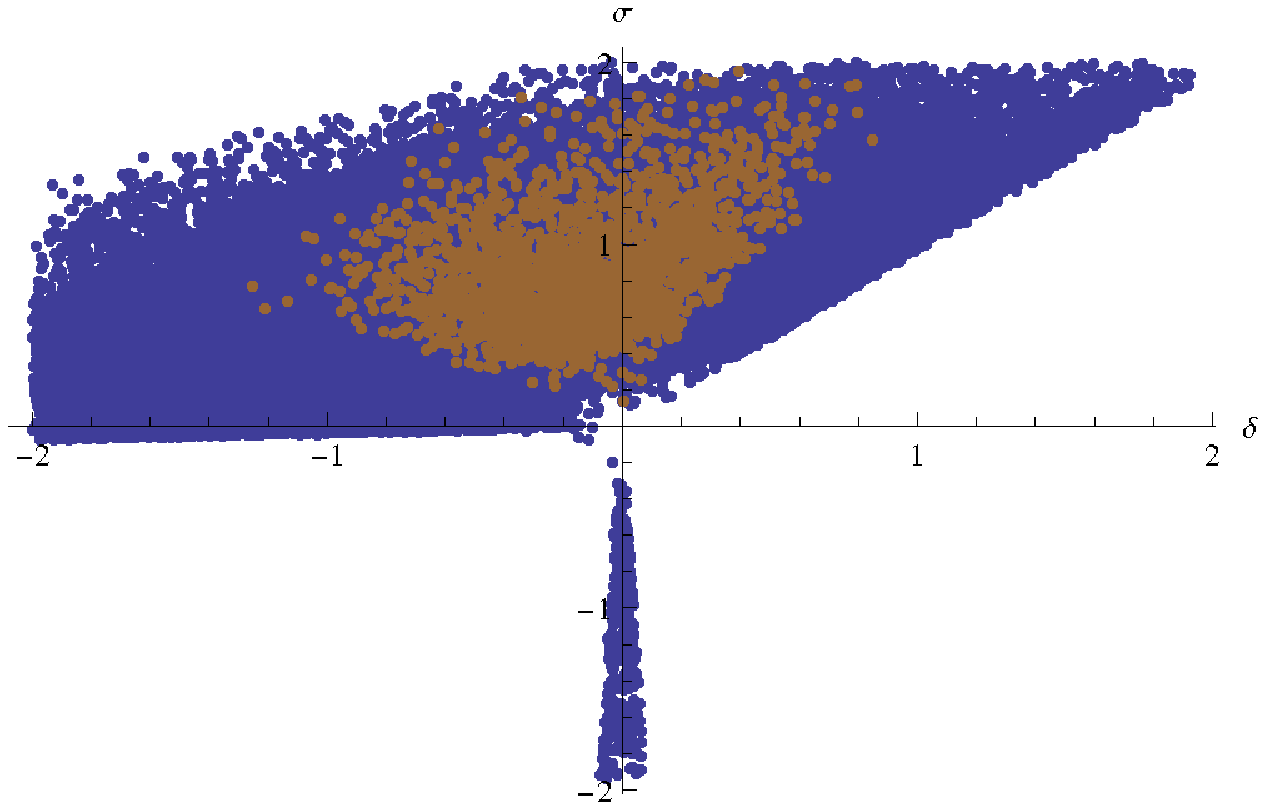}
\vskip 0.5 cm
\includegraphics[width=0.4\textwidth]{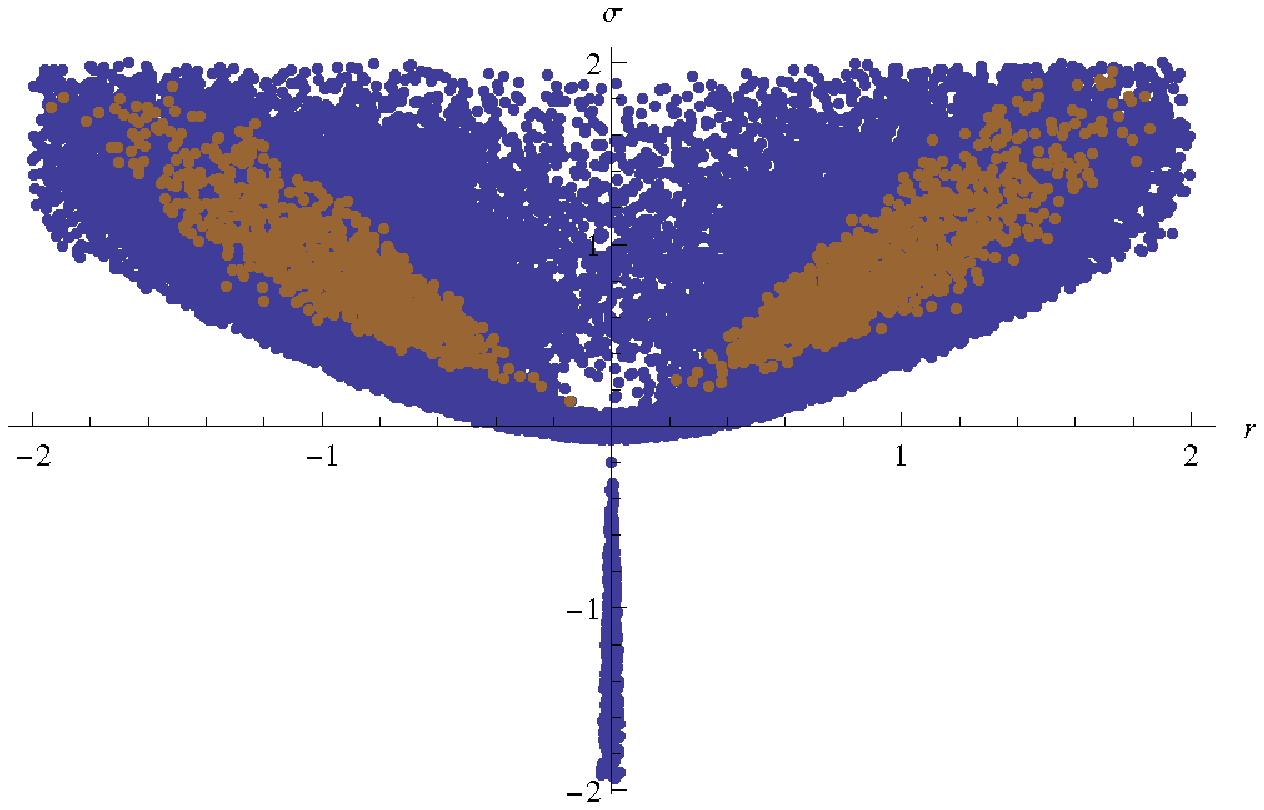}
\end{center}
\caption{Correlation between the free parameters $\delta$, $\sigma$, and $r$, after imposing the
experimental constrains on the neutrino mass matrix  in the case of normal hierarchy.
Blue points: squared mass constraints only. Brown points:  squared mass and mixing constraints
imposed.}
\label{fig:deltsig}
\end{figure}

\begin{figure}[t]
\begin{center}
\includegraphics[width=0.45\textwidth]{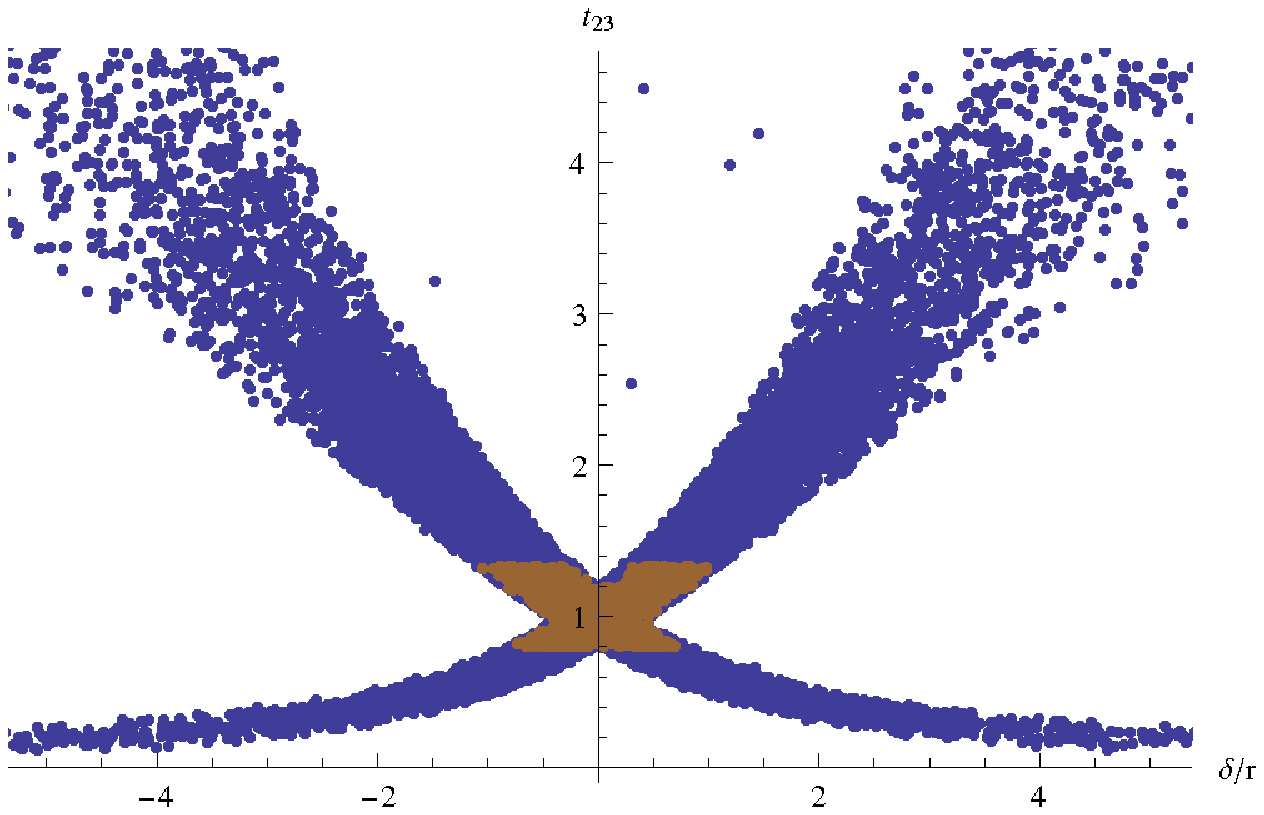} 
\vskip 0.5 cm
\includegraphics[width=0.45\textwidth]{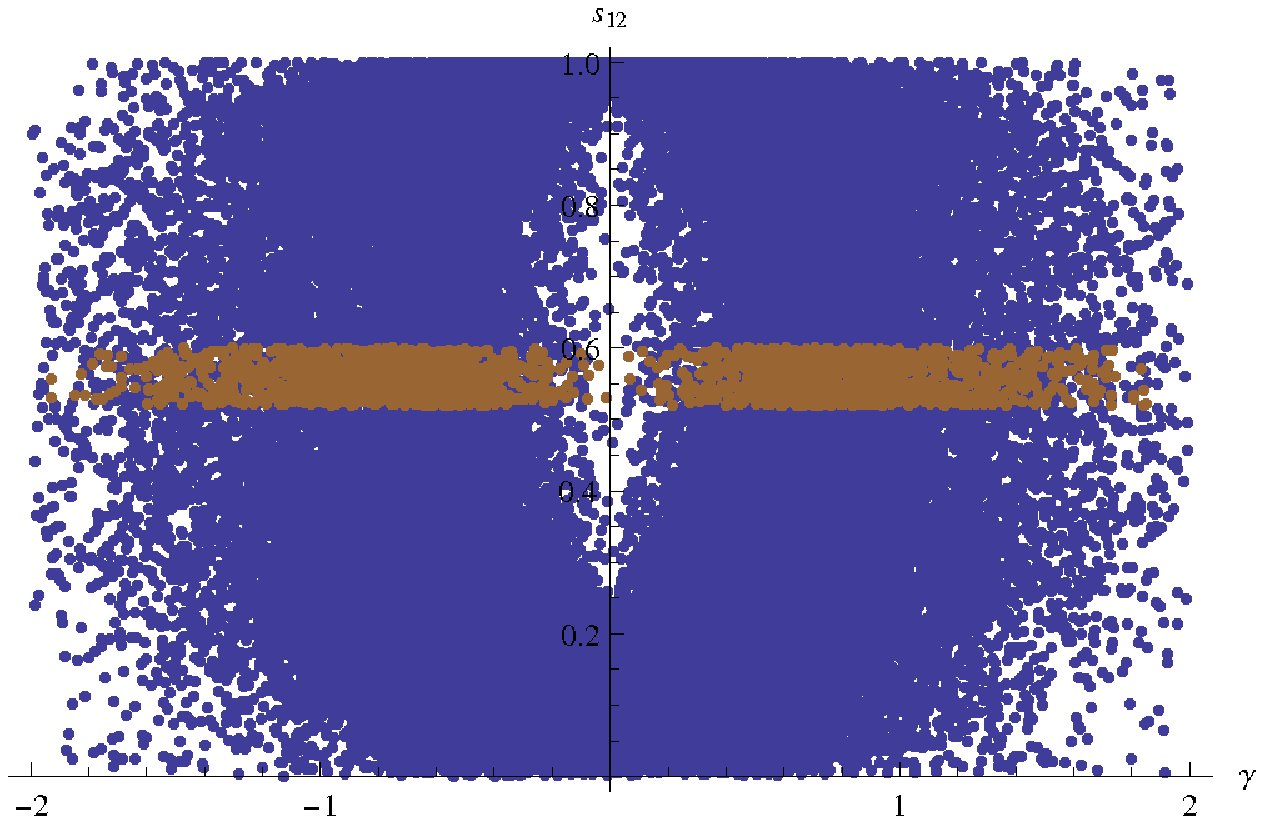}
\end{center}
\caption{Top: $t_{23}$ vs.~$\delta/r$.  Bottom: $s_{12}$ vs.~$\gamma$.  Notations as in Fig.~\ref{fig:deltsig}.}
\label{fig:angs}
\end{figure}

\begin{figure}[t]
\begin{center}
\includegraphics[width=0.45\textwidth]{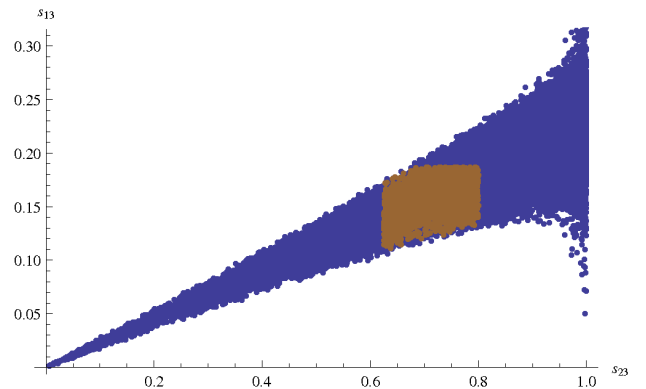} 
\end{center}
\caption{Correlation between $s_{13}$ and $s_{23}$ as expected in our framework. Notations as in Fig.~\ref{fig:deltsig}.}.
\label{fig:s13s23}
\end{figure}

In order to check the stability of the above conclusions we have performed a numerical scan allowing the four free real
parameters in  Eq.~(\ref{mat:mass}) to vary in the range $[-2,2]$. The results are summarized in Fig.~\ref{fig:deltsig}--\ref{fig:s13s23}.
In all plots the blue points are the allowed points after imposing the squared mass constraints only, while the points in brown are those 
where both squared mass and mixing constraints, as resulting from the global fit in Ref.~\cite{Schwetz:2011zk}, 
are satisfied.\footnote{Although the global fit of Ref.~\cite{Schwetz:2011zk} does not include the recent 
Daya-Bay result~\cite{An:2012eh}, the resulting value for $\theta_{13}$ turns out to be in good agreement with the 
direct determination in Eq.~(\ref{eq:s13exp}). 
 Similar results for all the neutrino parameters but for $\theta_{13}$ can be found also in Ref.~\cite{Fogli:2011qn}.}

The two plots in Fig.~\ref{fig:deltsig} illustrate the role of  $\sigma$, $\delta$, and $r$, in reproducing the mass spectrum.
For illustrative purposes, only the points giving rise to normal hierarchy are shown: the inverted case give rise to identical distributions
provided $\sigma\to-\sigma$ and $\delta\to-\delta$.
As can be seen from both plots, there is a wide range of values giving rise to the correct mass spectrum and no serious tuning of the
parameters is needed to explain the (modest) hierarchy between $|\Delta m_{\rm atm}|$ and  $\Delta m_{\rm sol}$. The latter emerge
naturally provided $\sigma$ is not too small. 

The results for $\theta_{13}$ and $\theta_{23}$ as a function of the corresponding most relevant free parameters are illustrated in Figure~\ref{fig:angs}
(again only the normal hierarchy case is explicitly shown).  On the top panel we show $t_{23}$ as a function of $\delta/r$: the two bands are those 
expected by the analytical expression in Eq.~(\ref{eq:t23}). As can be seen, we cannot claim to predict $t_{23} \approx1$, but this is a value 
perfectly allowed by our parameterization without particular fine tuning. On the contrary, very small or very large values of $t_{23}$
are disfavored after imposing the mass constraints.  On the bottom panel we show $s_{12}$   as a function of $\gamma$. As anticipated in the 
analytical discussion, $s_{12}$ is very difficult to be predicted (any value is essentially allowed). The only clear pattern emerging is
the need of a non-vanishing $\gamma$ in order to reproduce  the experimental value of $s_{12}$. 

In Figure~\ref{fig:s13s23} we show the correlation of $s_{13}$ and $s_{23}$, which is the only clear prediction of our decomposition,
as far as the mixing angles are concerned.  Also in this case we find a clean confirmation of what expected by means of the 
approximate analytical result in Eq.~(\ref{eq:t23}). The width of the band correspond to the uncertainty in the value of $s_e$,
that we have varied in the range $[0.19,~0.25]$.

\begin{figure}[t]
\begin{center}
\includegraphics[width=0.45\textwidth]{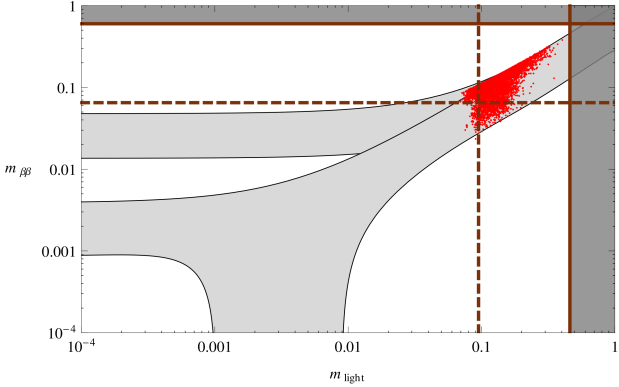} 
\end{center}
\caption{Correlation between the neutrino mass-matrix entry relevant to $0\nu\beta\beta$ experiments 
($m_{\beta\beta}$) and the lightest neutrino mass ($m_{\textrm{light}}$). The gray area is the area accessible 
from present oscillation experiments (at 3$\sigma$), the red points denote the prediction of our framework. The solid lines indicate the current bounds~\cite{Komatsu:2010fb};
the dashed lines provide an indication of the (near) future prospects~\cite{GomezCadenas:2010gs,Plaszczynski:2010sj}.}.
\label{fig:mbb}
\end{figure}

Finally, in Fig.~\ref{fig:mbb} we illustrate our predictions for the absolute values of neutrino masses. We find that typical values for the sum of neutrino masses lie around $0.2-1$~eV, in agreement with current bounds.\footnote{The WMAP bound of~\cite{Komatsu:2010fb} for the sum of neutrino masses varies between $1.3$~eV (WMAP-only) and $0.58$ (WMAP + Baryon Acoustic Oscillations + Hubble constant measurements)}  Interestingly, this means that neutrinoless double beta decay should be observed in the upcoming experiments, with the parameter $m_{\beta\beta}$ around $0.02-0.4$~eV. Current bounds for the matrix element, as well as the sensitivity of future experiments, are shown in Table~\ref{tab:nuless}.

\begin{table*}[t]
\begin{center}
{\small
\begin{tabular}{|c|rl||c|c|}
\hline
\multicolumn{3}{|c|}{Bounds} & \multicolumn{2}{|c|}{Prospects}\\
\hline
Experiment & Bound (eV), C.L. & & Experiment & Reach (eV)~\cite{GomezCadenas:2010gs} \\
\hline
KamLAND-Zen ($^{136}$Xe) & $<0.3-0.6$, 90\% & \cite{KamLANDZen:2012aa} & KamLAND-Zen ($^{136}$Xe) & $0.062$\\
CUORICINO ($^{130}$Te) & $<0.19-0.68$, 90\% & \cite{Arnaboldi:2008ds} & CUORE ($^{130}$Te) & $0.062$ \\
NEMO3 ($^{100}$Mo) & $<0.7-2.8$, 90\% & \cite{Arnold:2005rz} & NEXT ($^{136}$Xe) & $0.071$ \\
Heidelberg-Moscow ($^{76}$Ge) & $0.32\pm0.03$, 68\% & \cite{KlapdorKleingrothaus:2006ff} & EXO ($^{136}$Xe) & $0.072$ \\
\hline
\end{tabular}}
\end{center}
\caption{\label{tab:nuless}
Current bounds and prospects on $m_{\beta\beta}$. Intervals in bounds are due to uncertainty in the nuclear matrix elements. The ``reach" assume 
reference values for $\beta\beta$ isotope masses and a 10-year data taking period.}
\end{table*}

\section{The slepton sector and LFV}
\label{sec:pred.lfv}

\subsection{Slepton Structure}

Having identified the minimal set of spurions necessary to build the lepton Yukawa coupling and the neutrino mass matrix, 
we can now turn to study the consequences of this symmetry-breaking pattern in the slepton sector. 

Let's start from the $LL$ soft slepton mass matrix,
which transforms as  ${\bf 8}\oplus{\bf 1}$ under $U(3)_l$ and is invariant under $U(3)_e$.
The LN conserving spurions at our disposals are $Y^{(0)}$ and $\Delta \hat Y_e$, both transforming as 
bi-triplets of $U(3)_l \times U(3)_e$, 
and $X$, transforming as an ${\bf 8}$ of $U(3)_l$. Given the smallness of 
neutrino masses, we can safely neglect LN-conserving terms obtained by the contraction of two 
$m_\nu^{(0)}$ terms. Expanding to the first non-trivial  order in these spurions, the $LL$ soft mass matrix
assume the following form
\bea
 \tilde m^2_{LL}&=&\left(\begin{array}{c:c}
 (m^2_{L})_{\rm hh} & c_3 V^* \\ \hdashline
c_3 V^T & (m^2_{L})_{\rm 33}
\end{array}\right)\tilde m^2_{L}~, \no \\
 (m^2_{L})_{\rm hh} &=& I+c_3\Delta^*_L + c_4 \Delta Y_e^*\Delta Y_e^T~,   \no \\
 (m^2_{L})_{33}&=& 1+c_2 |y_\tau|^2 +c_3 x~,
\eea
with all constants being real and $\ord{1}$. Since 
$X$ is at most of $\ord{\epsilon}$ and $\Delta Y_e$ is at most of 
order $(y_\mu/y_\tau)$, we can approximate the above expression to 
\begin{equation}
\label{struc:LL} 
\tilde m^2_{LL}=\left(\begin{array}{ccc}
1 & c''_3\,\epsilon^2 & 0 \\
c^{\prime\prime *}_3\,\epsilon^2 & 1 + c_3\,\epsilon & c'_3\,\epsilon \\
0 & c^{\prime *}_3\,\epsilon & 1+c_2 |y_\tau|^2
\end{array}\right)\tilde m^2_{L}~,
\end{equation}
where we have distinguished $c_3$, $c'_3$ and $c''_3$ due to the possibility of additional $\ord{1}$ factors from the spurions themselves. 
With this definition, $c'_3$ and $c''_3$ are complex. In principle, these parameters are related to the parameters appearing in the neutrino 
mass matrix by 
\begin{align}
 {\rm Re} (c'_3)= \frac{1}{2}\frac{r}{\sigma-\delta}c_3~, & & {\rm Re} (c''_3)= \frac{1}{2}\frac{\gamma}{\sigma-\delta}c_3~.
\end{align}
However, we have explicitly checked that  these relations do not provide very stringent constraints. For this reason, 
in the numerical analysis we have treated $c_3$, $c'_3$, and $c''_3$ as independent free parameters. 

In the sfermion sector, the main difference between the $U(3)^5$ set-up we are considering, and that based on a $U(2)^5$ symmetry, lies on the fact that in the latter case one can naturally have sfermions of the first two families considerably heavier than those of the third family~\cite{Barbieri:2011ci}. As discussed in the introduction, this possibility is attractive due to the lack of experimental signals for supersymmetry at LHC and, at the same time, the need of relatively light  squarks of the third generation in order to stabilize the Higgs sector (hierarchy problem). 
In the $U(3)^5$ set-up we can also have third-family sfermions substantially lighter than those of the first two generations. However, this can happen only at the price of some fine-tuning of the  symmetry-breaking terms. In the case of 
$\tilde m^2_{LL}$, this happens if  $1+c_2 |y_\tau|^2 \ll 1 $. However, it is worth to stress that in the slepton sector the requirement of a sizable mass 
splitting among the families is less motivated: the sleptons play a minor role in the hierarchy problem and there are no stringent direct experimental bounds on any of the slepton families. 

The $RR$ soft slepton mass matrix transforms as  ${\bf 8}\oplus{\bf 1}$ under $U(3)_e$ and is invariant under $U(3)_l$.
Proceeding similarly to the $LL$ case we find 
\bea
\tilde m^2_{RR}&=&\left(\begin{array}{c:c}
 (m^2_{RR})_{\rm hh} & \Delta Y_e^T V^* y_\tau^* \\ \hdashline
 y_\tau V^T\Delta Y_e^* & (\tilde m^2_{RR})_{\rm 33} 
\end{array}\right) 
 \tilde m^2_{e_R^c}~,  \no \\
(m^2_{RR})_{\rm hh} & = & I + c_4 \Delta Y_e^T\Delta Y_e^* + c_5 \Delta Y_e^T\Delta_L^*\Delta Y_e^*~,  \no \\
 (m^2_{RR})_{\rm 33} &=& 1+y_\tau^*y_\tau(c_2+c_3 x)~.  
\eea
Here all off-diagonal terms are heavily suppressed by the first and second generation 
Yukawa couplings and, to a good approximation, can be neglected.

Finally, let's consider the trilinear soft-breaking term $A_e$,
responsible for the  $LR$ entries in the slepton mass matrices. 
The symmetry breaking structure of $A_e$ is identical to that of the Yukawas, 
albeit with different $\ord{1}$ factors:
\begin{equation}
 A_e=\left(\begin{array}{c:c}
 a_1 \Delta Y_e & a_2 V \\ \hdashline
 0 & a_3
\end{array}\right)y_\tau A_0~.
\end{equation}
Here the $a_i$ are complex $\ord{1}$ parameters. When diagonalizing the charged-lepton 
Yukawa we find
\begin{equation}
 (A_e)^Y \approx \left(\begin{array}{ccc}
 a_1 \ell_1 & 0 & (a_2-a_3) s_e\,e^{i\alpha_e} \epsilon \\
0 & a_1 \ell_2 & (a_2-a_3) c_e \epsilon \\
0 & 0 & a_3
\end{array}\right)y_\tau A_0~,
\end{equation}
where $\ell_1=(y_e/y_\tau)$ and $\ell_2=(y_\mu/y_\tau)$. This implies a negligible $LR$ contribution to the $1$--$2$ sector, and suppressed contributions to the $1$--$2,3$ sectors.

\subsection{Lepton Flavor Violation}

\begin{table}[t]
\renewcommand{\arraystretch}{1.3}
 \begin{center}
\begin{tabular}{|c|cl|rl|}
\hline
Channel & Bound (90\% C.L.)& & Prospects & \\
\hline
${\BR}(\mu\to e\gamma)$ & $<2.4\times10^{-12}$ & \cite{Adam:2011ch} & $10^{-13}$ & \\
${\BR}(\tau\to e\gamma)$ & $<3.3\times10^{-8}$ & \cite{Aubert:2009ag} & $10^{-9}$ &\\
${\BR}(\tau\to \mu\gamma)$ & $<4.4\times10^{-8}$ & \cite{Aubert:2009ag} & $10^{-9}$ & \\
\hline
 \end{tabular}
 \end{center}
\caption{Bounds and prospects for LFV searches.}
\label{tab:LFV}
\end{table}

Given the structure of the soft-breaking terms illustrated above, the leading contributions
to LFV processes are induced by LL terms.

Inspired by the symmetry-breaking patter of the squark sector analyzed in Ref.~\cite{Barbieri:2011ci},
and in order to to simplify the discussion, 
we start analysing the case where the third generation of 
sleptons is substantially lighter that the first two. In other words, we assume the existence 
of an approximate cancellation in the $(3,3)$ element of  the $LL$ slepton mass matrix. Under this assumption, 
the leading contributions to LFV processes are dominated by the exchange of third-family 
sleptons. 

Before analyzing the predictions of LFV rates by means of a numerical scan of the parameter space, we draw a few 
analytical considerations. In the limit where we assume the dominance of  chargino contributions
(as expected because of the larger coupling compared to neutralinos), we only need to analyze 
the LL mass matrix of Eq.~(\ref{struc:LL}). This is diagonalized by~\cite{Barbieri:2011ci}:
\begin{equation}
W^e_L = \left(\begin{array}{ccc}
 c_e &  s_e  e^{-i\alpha_e}  & -s_e s_L^e e^{i\gamma} e^{-i\alpha_e}  \\
-s_e e^{i\alpha_e}  &  c_e & -c_e s_L^e e^{i\gamma}   \\
  0  &  s^e_L e^{-i\gamma} & 1 \\
\end{array}\right)~,
\end{equation}
where 
\begin{equation}
s^e_L\,e^{i\gamma}=s_\tau e^{-i\phi_\tau}+ c'_3 =O(\epsilon)~.
\end{equation}
The relevant mixing terms are then 
\begin{eqnarray}
 \mathcal{R}^{\tilde\nu}_{13} &=& -s_e\,s^e_L\,e^{i(\gamma-\alpha_e)}~, \\
 \mathcal{R}^{\tilde\nu}_{23} &=& -c_e\,s^e_L\,e^{i\gamma}~, \\
 \mathcal{R}^{\tilde\nu}_{33} &=& 1~.
\end{eqnarray}
This allows us to make the approximate predictions:
\begin{eqnarray}
\label{rat:megtmg} 
\left(\frac{\BR(\mu\to e \gamma)}{\BR(\tau\to\mu\gamma)}\right)^{\chi^\pm} &\approx&
 \left(\frac{m_\mu}{m_\tau}\right)^5\frac{\Gamma_\tau}{\Gamma_\mu}
\left|\frac{\mathcal{R}^{\tilde\nu}_{23}\mathcal{R}^{\tilde\nu *}_{13}}
{\mathcal{R}^{\tilde\nu}_{33}\mathcal{R}^{\tilde\nu *}_{23}}\right|^2 \no\\
&\approx& 5.1\,s_e^2\,s^{e\,2}_L~, \\ \no \\
\label{rat:tegtmg}
\left(\frac{\BR(\tau\to e \gamma)}{\BR(\tau\to\mu\gamma)}\right)^{\chi^\pm} &\approx&
\left|\frac{\mathcal{R}^{\tilde\nu}_{33}\mathcal{R}^{\tilde\nu *}_{13}}
{\mathcal{R}^{\tilde\nu}_{33}\mathcal{R}^{\tilde\nu *}_{23}}\right|^2 \approx s_e^2~,
\end{eqnarray}
which turn out to be good approximations to the full results in the limit where 
third generation of sleptons is light.

In our numerical simulation, we include both chargino- and neutralino-mediated contributions. We perform a complete diagonalization 
of the full $6\times6$ slepton mass matrix and the $3\times3$ sneutrino mass matrix.\footnote{~In the diagonalization process we discard results 
with tachyonic sleptons or charged LSPs. We also take into account the approximate LEP bounds on chargino, stau and sneutrino masses~\cite{LEP}.}
 We take the $(3,3)$ and $(6,6)$ elements in the range $(200~{\rm GeV})^2-(1000~{\rm GeV})^2$, while we 
 assume values between $5^2$ and $100^2$ times heavier for the other mass eigenvalues. The $A_0$ parameter is assumed to be proportional to the heavy sfermion mass with a proportionality constant in the range $[-3,3]$. The chargino soft mass is fixed to $M_2=500$~GeV, 
and we use gaugino unification arguments to set $M_1=0.5M_2$. We also fix $\tan\beta=10$, and $\mu=600$~GeV.

\begin{figure}[t]
\begin{center}
\includegraphics[width=0.45\textwidth]{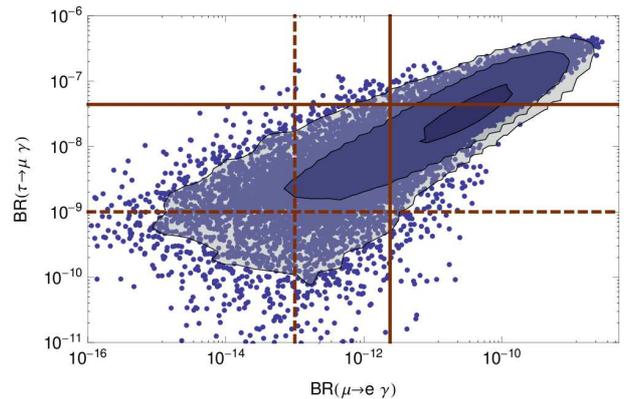}
\end{center}
\caption{Correlation between $\BR(\tau\to\mu\gamma)$ and $\BR(\mu\to e\gamma)$ in the case of 
hierarchical slepton mass 
spectrum ($\tilde m^2_{l_3} \ll \tilde m^2_{l_{1,2}}$).  See text for more details.} 
\label{fig:lfv}
\end{figure}
\begin{figure}[t]
\begin{center}
\includegraphics[width=0.45\textwidth]{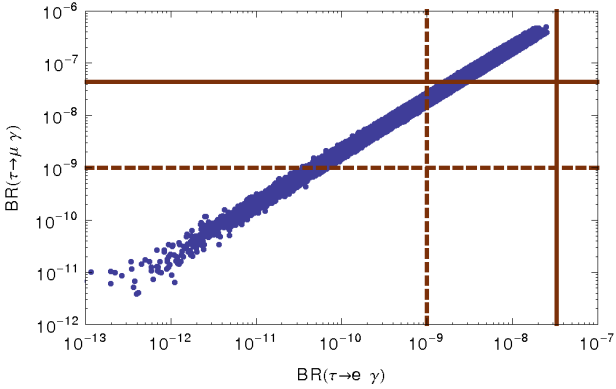}
\end{center}
\caption{Correlation between $\BR(\tau\to\mu\gamma)$ and $\BR(\tau\to e\gamma)$. See text for more details.} 
\label{fig:lfv2}
\end{figure}
\begin{figure}[t]
\begin{center}
\includegraphics[width=0.45\textwidth]{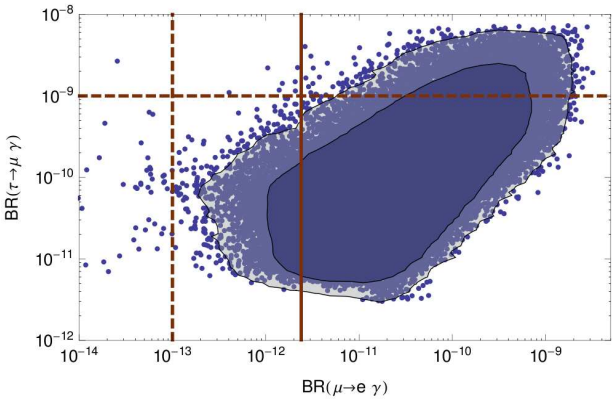}
\end{center}
\caption{Correlation between $\BR(\tau\to\mu\gamma)$ and $\BR(\mu\to e\gamma)$
in the case of almost degenerate slepton mass spectrum. See text for more details.} 
\label{fig:lfv3}
\end{figure}

The results of this numerical analysis are shown in Figure~\ref{fig:lfv}--\ref{fig:lfv2}. In Figure~\ref{fig:lfv} we show the correlation between $\BR(\tau\to\mu\gamma)$ and $\BR(\mu\to e\gamma)$, while on the bottom panel we show the correlation of the former with $\BR(\tau\to e\gamma)$. We show the current bounds for each branching ratio with solid brown lines, while the expected sensitivity of the relevant experiment (MEG for $\mu\to e\gamma$, Belle II and SuperB for $\tau\to \ell\gamma$) is shown using dashed brown lines (see Table~\ref{tab:LFV}).
The contours on the left panel give an idea of the density of points, where each outer contour represents a density value one order of magnitude smaller than the respective inner contour.

Figure~\ref{fig:lfv} shows that, although a small part of the parameter space is ruled out already, there exist a significant number of points that can be probed by $\mu\to e\gamma$, $\tau\to\mu\gamma$ and possibly also  $\mu\to e$ conversion experiments in the near future. It is interesting to note that current and future sensitivities of $\mu\to e\gamma$ and  $\tau\to\mu\gamma$ are quite comparable in constraining the model, even if the experimental sensitivity on 
$\tau\to\mu\gamma$ is much weaker. The fact that $\tau\to\mu\gamma$ has such an important role can easily be understood from Eq.~(\ref{rat:megtmg}), where it is clear that $\mu\to e\gamma$ receives an additional suppression due to $s_e^2$. Similar conclusions have recently been reached also in Ref.~\cite{Barbieri:2012uh}. On the other hand, the correlation between $\mu\to e\gamma$ and $\tau\to\mu\gamma$ in Figure~\ref{fig:lfv}  is quite different with respect what expected in various models of Minimal LFV~\cite{Cirigliano:2005ck}, where there is no connections between quark and lepton flavor structures.

Figure~\ref{fig:lfv2} shows that $\tau\to e\gamma$ does not provide additional bounds on the model,
as most points that can be probed by this decay mode are already ruled out by $\tau\to\mu\gamma$. Still, the Figure shows a very strong correlation, as expected from Eq.~(\ref{rat:tegtmg}). This correlation could provide a very significant test of the model if it could be verified experimentally.

Finally, in order to test how these conclusions are modified if the slepton spectrum is not hierarchal, 
we have performed a independent scan without assuming a cancellation in the  $(3,3)$ and $(6,6)$ 
entries of the slepton mass matrix. In particular, we vary all the diagonal entries 
in the range $(1000~{\rm GeV})^2-(2000~{\rm GeV})^2$, while keeping all the other 
parameters (gaugino and chargino masses) fixed as in the previous scan. 
The result of this second numerical analysis  are shown in Figure~\ref{fig:lfv3}.
As can be seen, in this case the correlation between $\mu\to e\gamma$ and  $\tau\to\mu\gamma$
is quite different: the contribution of the sleptons of the first two families 
is not negligible in  $\mu\to e\gamma$ and,  as a consequence, the approximate relation Eq.~(\ref{rat:megtmg})
is no longer valid. In this framework, the recent MEG bound~\cite{Adam:2011ch} on $\mu\to e\gamma$ provide a very severe
constraint. In particular, it rules out the possibility of visible effects in the  $\tau\to\mu\gamma$ case. 
The correlation between $\tau\to\mu\gamma$ and $\tau\to e \gamma$  is not modified with 
respect to Figure~\ref{fig:lfv2}, but both modes are beyond the experimental reach after 
imposing the $\mu\to e\gamma$ bound.

\section{Conclusions}

We have proposed an ansatz for the neutrino mass matrix and the charged lepton Yukawa coupling
based on a minimal breaking of the  $U(3)^5$ flavor symmetry, consistent with the $U(2)^3$ 
breaking pattern of the quark Yukawa couplings discussed in Ref~\cite{Barbieri:2011ci}. 
The key hypothesis that allows us to relate the non-hiearchical neutrino sector to the Yukawa sector
is the assumption of a two-step breaking structure in the neutrino case:  a leading breaking of the maximal flavor symmetry, 
$U(3)_l \times U(3)_e$, giving rise to a fully degenerate neutrino spectrum, 
followed by a sub-leading hierarchical breaking similar to the one occurring in the Yukawa sector. 
According to this hypothesis, the large 2-3 mixing in the neutrino sector arises as a small perturbation 
of an approximately degenerate spectrum. On the other hand, the ratio between $\theta_{13}$ and $\theta_{23}$ 
is predicted to be of the order of the Cabibbo angle, similarly to the quark sector, in good agreement   
with the recent DayaBay result. 

As we have shown,  our framework is able to reproduce all the neutrino oscillation parameters without particular tuning of the free parameters. 
The neutrino masses are predicted to be almost degenerate: the sum of all the eigenvalues is expected to be around $0.2-1$~eV,
close to the present cosmological bounds, and the  $0\nu\beta\beta$ parameter $m_{\beta\beta}$ is expected in the range $0.02-0.4$~eV, observable in next generation of experiments. 

Our framework can naturally be implemented in supersymmetric extensions of the SM and, more explicitly, within  the well-motivated 
set-up with heavy masses for the first two generations of squarks. We have analyzed the consequences 
of this flavor-symmetry breaking ansatz in the supersymmetric case, assuming a split family spectrum also in the slepton sector.
The model can satisfy the existing constraints on LFV in charged leptons, with the most significant bounds coming from 
$\mu\to e\gamma$ and $\tau\to\mu\gamma$. For third-generation sleptons masses below 1 TeV both decay modes are expected to 
be within  the reach of future experimental searches.

\section*{Acknowledgements}
This work was supported by the EU ERC Advanced Grant FLAVOUR (267104),
and by MIUR under contract 2008\-XM9HLM.
G.I. acknowledges the support of the Technische Universit\"at M\"unchen -- Institute for Advanced
Study, funded by the German Excellence Initiative.

\bibliographystyle{utphys}

\end{document}